\documentclass[fleqn,usenatbib]{mnras}

\usepackage[T1]{fontenc}

\usepackage{graphicx}	
\usepackage{amsmath}	
\usepackage{amssymb}	
\usepackage{epstopdf}
\usepackage{wrapfig}
\usepackage{float}
\usepackage{adjustbox}
\usepackage{rotating}
\usepackage{ltablex} 
\usepackage{supertabular,booktabs}
\usepackage{longtable}
\usepackage{graphicx}
\usepackage{amsmath,amssymb}
\usepackage{amsfonts}

\usepackage{multirow}
\usepackage[normalem]{ulem}
\useunder{\uline}{\ul}{}
\newcommand{\be}{\begin{equation}} 
\newcommand{\ee}{\end{equation}}

\usepackage[acronym]{glossaries}
\newacronym{msma}{MSMA}{Multi-Stage Memetic Algorithm}
\newacronym{ecblof}{ECBLOF}{Enhanced Cluster Based Local Outlier Factor}
\newacronym{msmvmca}{MSMVMCA}{Multi-Stage Multi-Version Memetic Clustering Algorithm}
\newacronym{cblof}{CBLOF}{Cluster-Based Local Outlier Factor}
\newacronym{lc}{LC}{Large Cluster}
\newacronym{sc}{SC}{Small Cluster}
\newacronym{at}{AT}{Anomaly Tree}
\newacronym{msmbtai}{MSMBTAI}{Multi-Stage Memetic Binary Tree Anomaly Identifier}
\newacronym{iforest}{iForest}{Isolation Forest}
\newacronym{lof}{LOF}{Local Outlier Factor}

\usepackage{xcolor}
\definecolor{lb}{RGB}{44, 139, 183}

\usepackage[linesnumbered,ruled,vlined]{algorithm2e}
\SetKwComment{Comment}{$\triangleright$\ }{}
\SetCommentSty{itshape}
\usepackage{nomencl}
\usepackage{mathtools}
\usepackage{makecell}

\newtheorem{example}{Example}
\def\be{\begin{equation}}
\def\ee{\end{equation}} 
\usepackage{textcomp} 
 
\title[Exoplanetary Habitability as Anomaly]{Postulating Exoplanetary Habitability via a Novel Anomaly Detection Method}
 
\author[Jyotirmoy Sarkar et al.]{
Jyotirmoy Sarkar,$^{1}$
Kartik Bhatia,$^{1}$
Snehanshu Saha,$^{2}$
Margarita Safonova$^{3}$
and Santonu Sarkar$^{1}$\\
$^{1}$BITS Pilani, K.~K.~Birla Goa Campus, Goa, India\\
$^{2}$CSIS and APPCAIR, BITS Pilani, K.~K.~Birla Goa Campus, Goa, India\\
$^{3}$Indian Institute of Astrophysics, Bangalore, India
}
 
\date{}

\pubyear{}

\begin{document}
\label{firstpage}
\pagerange{\pageref{firstpage}--\pageref{lastpage}}
\maketitle

\begin{abstract}
A profound shift in the study of cosmology came with the discovery of thousands of exoplanets and the possibility of the existence of billions of them in our Galaxy. The biggest goal in these searches is whether there are other life-harbouring planets. However, the question which of these detected planets are habitable, potentially-habitable, or maybe even inhabited, is still not answered. Some potentially habitable exoplanets have been hypothesized, but since Earth is the only known habitable planet, measures of habitability are necessarily determined with Earth as the reference. Several recent works introduced new habitability metrics based on optimization methods. Classification of potentially habitable exoplanets using supervised learning is another emerging area of study. However, both modeling and supervised learning approaches suffer from drawbacks. We propose an anomaly detection method, the Multi-Stage Memetic Algorithm (MSMA), to detect anomalies and extend it to an unsupervised clustering algorithm MSMVMCA to use it to detect potentially habitable exoplanets as anomalies. The algorithm is based on the postulate that Earth is an anomaly, with the possibility of existence of few other anomalies among thousands of data points. We describe an MSMA-based clustering approach with a novel distance function to detect habitable candidates as anomalies (including Earth). The results are cross-matched with the habitable exoplanet catalog (PHL-HEC) of the Planetary Habitability Laboratory (PHL) with both optimistic and conservative lists of potentially habitable exoplanets. 
\end{abstract}
\begin{keywords}
Anomaly Detection -- Machine Learning -- Unsupervised Learning -- Exoplanets -- Habitability
\end{keywords}
 
\section{Introduction}

Thousands of exoplanets have been discovered in last few decades, with implication that there are more planets than stars in our Galaxy. Ultimately, we are searching for habitable planets or, at least, for potentially habitable. By potentially habitable planets we understand those classes or types of planets whose properties indicate their ability to beget or sustain life, such as e.g. being a rocky planet in a habitable zone (HZ) of the host star allowing for a liquid water on the surface, and so on. By applying various Earth-based criteria, more than one potentially habitable exoplanets have indeed been hypothesized. The Habitable Exoplanets Catalog (HEC) maintained by the PHL\footnote{The Habitable Exoplanets Catalog (HEC) is an online database of potentially habitable planets maintained by the Planetary Habitability Laboratory @UPR Arecibo; available at http://phl.upr.edu/projects/habitable-exoplanets-catalog} describes a small fraction of all discovered planets as potentially habitable. These are divided into two categories: optimistic (24 planets) and conservative (36 planets) lists of potentially habitable exoplanets. Since the number of potentially habitable exoplanets is significantly less than the non-habitable ones\footnote{60 in the latest update of the catalog, October 5, 2020.}, these habitable candidates could be thought of as anomalies in a large pool of normal (non-habitable) instances. The idea of equating the habitability detection problem to an anomaly detection problem therefore deserves merit, and no method supporting this hypothesis exists currently in the literature. The idea resonates well with the fact that Earth is the only known habitable planet among several thousand detected, and there is a possibility that life is arbitrarily rare \citep{rarelife}. Backed by such physical observations, it is reasonable to postulate Earth as an anomaly, and hypothesize the potentially habitable planets as anomalies. This will allow the anomaly detection, i.e. habitable exoplanet detection problem, to be framed as an unsupervised machine learning (ML) approach via a novel clustering algorithm. Consequently, we can estimate the absolute prerequisites for habitability in order to set up a list of planets harbouring life indicators, using the examples from our own planet.

Our hypothesis is that this small fraction of planets are anomalous instances in a large pool of `non-habitable' exoplanets (a total of 4383 planets are confirmed so far, with 5912 candidates\footnote{NASA Exoplanet Catalog, May 3, 2021.}). With this large number of discovered exoplanets, it is imperative to examine these rare instances by characterizing all exoplanets in terms of planetary parameters, types, populations and, ultimately, the habitability potential. This needs the knowledge of multiple planetary parameters from observations which, in turn, demands hours of expensive telescope time. Therefore, we need to prioritize the planets to examine, i.e. develop a quick screening tool for evaluating habitability perspectives from observed properties.

\section{Habitability Quantification and Classification: Existing Approaches}\label{sec:RelatedWork}

Detecting potentially habitable exoplanets using machine learning tools have gained momentum in recent times \citep{old,new,ceesa}. This research suggests that the habitability can be viewed as probabilistic measure \citep{old}, and such approaches require optimization and classification methods. Two popular approaches to habitability estimation are discussed below.

\subsection{Metric-based quantification}

\citet{old} introduced a Cobb-Douglas Habitability Score -- a metric based on Cobb-Douglas habitability production function (CD-HPF), which computes the habitability score by using measured and estimated planetary parameters. \citet{ceesa} extended the CDHS model and proposed another quantitative metric for habitability -- CEESA, which considers orbital eccentricity in addition to the same features used by the CDHS. These metrics, based on optimization methods, use only four physical planetary parameters (mass, density, radius and surface temperature), while there may be a need to accommodate more features such as, for ex., eccentricity, or orbital separation. In \citet{limbach} it was proposed that low eccentricity favours multiple planetary systems which, in turn, favours habitability as it may stabilize the climate on a planet \citep{wang}. While it is acceptable to reason that many factors of life are dependent on temperature, habitability metrics should be assessed on more parameters than just the temperature. Factors such as escape velocity, eccentricity and many others, are important while determining whether a planet can be potentially habitable or not. For example, the brown dwarf WD~0806-661B has the surface temperature of $52^{\circ}-77^{\circ}$C \citep{Luhman}, which is in the habitable range (Earth average temperature is $\sim$15$^{\circ}$C), but the planet has a mass of 7 to 9 $M_{\rm Jup}$ --- too massive to be potentially habitable. Some researchers (e.g. \citet{tasker}) even suggest that it is impossible to compare habitability on different planets quantitatively based on a single habitability score or metric.

\subsection{Supervised learning}
\label{sec:2.2}

\citet{new} applied a statistical ML classification method (XGBoost), where the training data with multiple features are used to predict a target variable. The classification method is supervised and rely heavily on training labels. Classification strategies rely on class labels primarily based on surface temperature of the planets. Since training labels are based on measurement of surface temperature, this is often an issue as many of the discovered exoplanets don't possess measured surface temperature. As many of the class labels are processed based on estimated surface temperature, even high training and validation accuracies are not free from scrutiny. \citet{saha2019evolution} have shown recently that the classification accuracy drops sharply if features related to surface temperature are removed. \citet{mendez} discusses that the surface temperature-based bounds for habitability are approximate and there are species who could sustain life in warmer and cooler climates compared to the limits specified by these temperature boundaries for habitable or non-habitable classes. Surface temperature-based class labels cannot be entirely trusted. Therefore, an automated learning process for detecting potentially habitable exoplanets using Earth as a reference needs to be independent of class labels.

Anomalous instances are hard to detect, especially when the data are not labelled. The DeepAnT \citep{Munir} approach uses unlabeled data for training, but it is tightly coupled with time-series in the dataset. Like any other Deep Learning method, this approach requires retraining from scratch for a different kind of data type. Deep Learning-based anomaly detection methods are primarily supervised approaches. A clustering-based approach -- Cluster Centers -- has been proposed to identify anomalies \citep{Castellani}. It has been designed for a dataset that has very few labeled data. Isolation-based algorithms, such as \acrshort{iforest} and K-means-based Isolation Forest are the unsupervised approaches. \citet{Karczmarek} considered anomaly along with normal data with the aim to isolate anomalous instances from the rest of the data.

This makes a strong case for unsupervised approach to the habitability detection problem. Thus, we remove the class labels and steer clear of wrong annotation and explainability problems and proceed to solve the anomaly detection problem via clustering. An Anomaly is an instance that occurs rarely and may be present among a pool of normal data instances. We propose an efficient detection algorithm capable of flagging several anomalies, if present, in the data set. 

\subsection{Summary of Proposed Contributions:}

Toward the goal of detecting potentially habitable planets other than Earth i.e. anomalies, we contribute to the literature of unsupervised learning and propose:
\begin{itemize}
\item A novel clustering method, MSMVMCA, based on a novel multi-stage memetic algorithm (MSMA);
\item \acrfull{msmbtai};
\item \acrfull{ecblof}, a distance semi-metric;
\item Application of \acrfull{msmbtai} and \acrfull{ecblof} to detect anomalies i.e. potentially habitable exoplanets from data.
\end{itemize}

We will show that \acrshort{msmvmca} can use multiple fitness functions and therefore is better equipped to handle anomalies. The proposed clustering algorithm have been tested on benchmark data sets and have been compared with other methods. Since these data sets are not related to the exoplanet data, results are available on request. We have shown that \acrshort{msmvmca} outperforms benchmark methods on additional data sets.
 
\section{Multi-Stage Multi-Version Memetic Clustering Algorithm (MSMVMCA)}
\label{MSMVMCAsection}

Memetic algorithms are a class of metaheuristic algorithms, where an evolutionary approach is hybridized with the problem-specific information. The objective of the hybridization is to accelerate the discovery of good solutions, for which the EA (Evolutionary Algorithm) would have taken a long time to reach, or it was never possible to reach \citep{Aragn}. In \acrshort{msmvmca} (Algorithm \ref{algo:MSMVMCA}), the initial population is designed for supporting the clustering features. \acrshort{msmvmca} is based on the principles of \acrfull{msma}. The significant difference of \acrshort{msma} with other typical memetic algorithms is that \acrshort{msma} supports multiple stages. Though we have implemented two stages in the manuscript, it is possible to increase the number of stages. Typically, a memetic algorithm employs a single crossover, mutation rate, and a single fitness function. However, \acrshort{msma} can apply multiple strategies, where different mutation rate, crossover rate, and fitness functions are used. A set of crossover rate, mutation rate, and fitness function forms a single stage. Mult-stage allows multiple mutations, crossover rates, and fitness functions in different stages.  

\subsection{Various Memetic Definitions}

\subsubsection{Mutation}

The mutation operator of the memetic algorithm is analogous to the biological mutation. Mutation alters one or more gene values in the chromosome from its initial state. The rate of mutation differs from problem to problem. Say, a chromosome length is $10$, and the mutation rate is  $0.1$. This implies $1 (0.1*10)$ gene is eligible for mutation. The individual solution is represented as a chromosome. The gene is selected randomly for mutation. For \acrshort{msmvmca}, a typical solution or chromosome is $0 1 0 1 1 0 1$ and the probable values are $0,1$ (two clustering problems). The individual value of the chromosome is considered a gene. If the fourth element is selected for mutation, the value will be flipped $0$. Then the new solution looks like $0 1 0 0 1 0 1$.
\subsubsection{Crossover}
The crossover functionality of a memetic algorithm is similar to the biological crossover. It is also called recombination. Here, two solutions are considered as parents, and they exchange their genetic information to produce a new offspring or solution.
For example, consider $0 0 1 1 1 1$ and $0 1 1 1 0 0$ as the parents and the crossover rate as $0.5$. Then $3$ genes ($0.5 *$size of the solution) from a parent will be exchanged to the other parent. The new solutions will be $0 1 1 1 1 1$ and $0 0 1 1 0 0$.
\subsubsection{Fitness Function}
Fitness function evaluates the fitness of the new solution. The solutions which have higher fitness are eligible for survival. Every problem uses different fitness function specific to that domain.

\subsection{MSMA: Operators Overview}

MSMA describes four operators ($\Omega_i$ for $i = {1, 2, 3, 4}$) based on Genetic heuristic. 
 The operators $\Omega_1$ and $\Omega_2$
are used in Stage 1, and $\Omega_3$ and $\Omega_4$ are used in Stage 2. \\
\noindent \textbf{Stage 1 Breeding Strategy ($\Omega_1$):}
\(\Omega_1\) operates on the initial multi-version gene pool. For each pair of versions (chromosome), say father and mother version, select a random index t (split location) in [1; k].\\
We transfer the class values (genes) $x_1$ to $x_t$ from mother version and the genes from $x_t+1$ to $x_k$ from father version to child1. The remaining class values from mother and father are transported to their respective positions in child2. Thus, two children are created from the versions (chromosomes) of father and mother. \\
\noindent \textbf{Stage 1 Mutation Strategy ($\Omega_2$):} A predefined ratio of class value (gene) in child version is replaced with the random value to produce mutant. Here a different approach has been taken for mutation, and it is called a selective mutation. For a fixed number of times, it tries to find a mutant which improves the fitness value of the chromosome (version) in comparison to the parent. If it finds, then it replaces the version with the newly found mutant.\\
\noindent \textbf{Stage 2 Breeding Strategy $\Omega_3$:}
$\Omega_3$ operates similarly to $\Omega_1$, though the transfer ratio is not the same.\\
\noindent \textbf{Stage 2 Mutation Strategy ($\Omega_4$):} Similar to operator $\Omega_2$, a predefined ratio of the total class values (gene) in a version is selected and replaced by randomly generated class value. 
The predefined ratio for the mutation in $\omega_4$ can be different from $\Omega_2$. The same selected mutation strategy has been applied here, like operator $\Omega_2$.

\begin{algorithm}
\caption{ MSMVMCA(D,v,$s_1$,$s_2$,$n_1$,$n_2$,$F_t$)}
\label{algo:MSMVMCA}
\SetAlgoLined
\KwResult{version with the best fitness value }
\KwIn{D -- input data, $v$ -- the number of versions, $s_1$ -- the proportion of the population in stage1 becoming parents, $s_2$ -- the proportion of the population in stage2 becoming parents, $n_1$ is the max stage1 iteration number, $n_2$ is the max stage2 iteration number, $F_t$ is the fitness threshold value.}
population $\leftarrow$ random new multi-version population-based on D and $v$;\\
$outer_{itr} \leftarrow 0$;

\While{$outer_{itr} \leq n_1$ or isStoppingCriteraMeet()}{
Initialize fitnessArray \\

\For{each version $v$ in  population}{fitnessArray$\leftarrow$ Silhouette(v,D) }

Sort(fitnessArray,population)
parents $\leftarrow$ top $s_1$ proportion of population\\
survivors $\leftarrow$ top ($1-s_1$) proportion of population \\
children $\leftarrow \Omega_1$(parents) \\
mutants $\leftarrow \Omega_2$(children) \\
population $\leftarrow$ survivors $\cup$mutants \\

$inner_{itr} \leftarrow 0$ 

\While{$inner_{itr}\leq n_2$ or  isStoppingCriteraMeet()}{
Initialize innerFitnessArray \\
\For{each version $v$ in  population}{
innerFitnessArray$\leftarrow$ DaviesBouldin(v,D) 
}
Sort(innerFitnessArray,population) 
parents $\leftarrow$ top $s_2$ proportion of population;\\
survivors $\leftarrow$ top ($1-s_2$) proportion of population \\
children $\leftarrow \Omega_3$(parents) \\
mutants $\leftarrow \Omega_4$(children) \\
population $\leftarrow$ survivors$\cup$mutants \\
$inner_{itr}\leftarrow inner_{itr}+1$
}

\If{Max-fitness(population)$\geq$ $F_t$}{
Break
}
$outer_{itr}\leftarrow outer_{itr}+1$
}
\Return{version $\leftarrow$ Max-fitness(population)}
\end{algorithm}

\noindent \textbf{Population initialization:}
The initial population is a multi-version gene pool described in Table \ref{table:initpopu}. This is generally called Integer Label-based Encoding \citep{Murthy}. Each individual is a vector of size $N$, where $N$ is the number of data points. Each position in the vector takes a value from 0 to $K-1$, where $K$ is the total number of clusters.

\begin{table}
\centering
\caption{The initial population of the \acrshort{msmvmca} algorithm: where the rows are versions and columns represents the data points. In a different version, the data point is assigned to a different class. $M[0][2]=2$ means in the first version; the third data point is assigned to the third class. (Indexing starts from $0$, and $M$ represents population matrix)}
\label{table:initpopu}
\begin{tabular}{|c| c| c| c| c| c| c| c| c| c| c|} 
\hline 
Version No&1& 2 & 3&4&5& 6 & 7&8&9&10\\ 
\hline
V$_1$ & 0 & 1 & 2 & 2& 0& 1 &  1&2 &0 & 0\\ 
V$_2$ & 0 & 0 &1&2&2&2&1&0&0&1\\
V$_3$ & 1 &  1& 0& 0& 2&  1& 1& 1 &0 & 1\\
V$_4$ &0 &2 & 2 & 1& 1&2 &0 & 1 &0 &0\\
\hline
\end{tabular}
\end{table}

\noindent \textbf{Characteristics of \acrshort{msmvmca}:}\label{MSMVMCA}
The proposed algorithm is driven by the clustering purity metric, which acts as a fitness function. During the clustering process, using the operators (\(\Omega_1\) to $\Omega_4$), it attempts to find the best possible version, which produces better fitness value. This is equivalent to hybridization of EA by introducing clustering purity metric as the fitness function, and the initial gene pool is designed to support various versions of clustering solutions. \acrshort{msmvmca} employs multiple crossover and mutation rates, which is not present in a typical EA supporting a single fitness function.  
Once \acrshort{msmvmca} terminates,  we gather a set of versions, and each version is a probable clustering solution. Though we consider the version having the highest fitness value, all the versions with fitness values greater than a predefined threshold may participate in producing the final solution via majority voting. Data points are assigned to a cluster supported by the majority of the versions. Say, $V_1$, $V_2$ and $V_3$ have fitness value higher than the threshold. $V_3$ suggests that the data point1 belongs to cluster 1, but $V_1$ and $V_2$ suggest the data point is close to cluster 0. Therefore, by majority voting, the data point1 is assigned to cluster 0. \acrshort{msmvmca} qualifies as a clustering algorithm and has been applied on multiple benchmark datasets (See Appendix~\ref{appen-msmvmca} for more details).

\subsection{Multi-Stage Memetic Binary Tree Anomaly Identifier (MSMBTAI)}\label{sec:msmbtai}

\begin{figure}
  \centering
      \includegraphics[width=.5\textwidth]{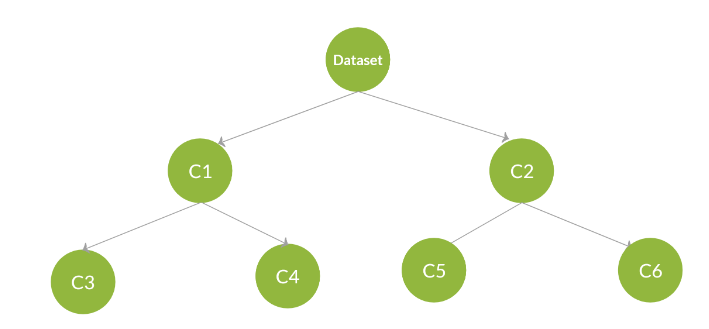}
\caption{This figure demonstrates the recursive nature and a sample tree structure of the MSMBTAI.}
\label{matreelatest}
\centering
 \end{figure}
\noindent In order to evaluate the efficacy of a clustering based anomaly detector, the \acrfull{cblof} metric proposed by \citep{Deng} has been used extensively. We propose the  \textbf{Enhanced Cluster Based Local Outlier Factor (\acrshort{ecblof})}, a modified CBLOF metric, where we avoid the multiplication with the cluster size since such a score can bias towards the large clusters. The following example explains the rationale.
\begin{example}
Suppose that a small cluster $C_1$ has only one data point $a$ which is an anomaly. Consider another large cluster $C_2$ with a non-anomalous data point $b\in C_2$. Suppose that the distance between $a$ and centroid of $C_2$ cluster is 10 ($d(a,C_2)=10$). Since $C_2$ is the nearest larger cluster, as per the definition of the \acrshort{cblof}, the anomaly score of $a=10\times1$, whereas that of $b=1*|C_2|$ (since $b\in C_2$, its distance from the centroid of $C_2$ will be 1). If $|C_2|>10$, then $a$ will not be identified as an anomaly based on the \acrshort{cblof} score and a normal data point $b$ will be considered as anomaly because of the higher anomaly score\footnote{Anomaly detection is based on designing efficient clustering algorithms. We have shown that, \acrshort{msmvmca}, our proposed anomaly detector, is an efficient clustering method. See Appendix A for details.}. 
\end{example}
The modified metric called \acrshort{ecblof} is defined below:
Let $D$ be the dataset, partitioned into $n$ clusters $C_1,\cdots,C_i\cdots,C_n$. We categorize the set of clusters into two, namely \acrfull{sc} and \acrfull{lc}. For a predefined threshold $\alpha$, $C_i\in SC$ if $|C_i|<\alpha|D|$, otherwise it is in $LC$. 
For a data point $p\in D$, we define $ECBLOF(p)$ as
\begin{align}
     &\begin{aligned}\label{scecblof}
     ECBLOF(p)  =\begin{cases}
             \quad min(d(p,C_j))\ if\ C_i \in SC\\ \quad where \quad p\in C_i \wedge C_j \in LC\\ \quad \forall j=1\cdots b
          \end{cases}
     \end{aligned}\\
     &\begin{aligned}\label{lcecblof}
     ECBLOF(p) = d(p,C_i)\ if\ C_i\in LC\ where\ p\in C_i
     \end{aligned}
\end{align}
\textbf{\textit{$d(p, C_j)$ computes the distance between p and the cluster center of $C_j$.}} 
\subsection*{Anomaly Detection using MSMBTAI:}
The entire anomaly detection process is divided into two phases. First phase is the construction of the \acrfull{at} using the proposed \acrshort{msmvmca}. In the second phase, MSMBTAI tries to find the probable anomalies using the \acrshort{ecblof} metric. Algorithm \ref{BATAlgo} constructs the \acrshort{at} and a similar tree structure of the \acrshort{at} is demonstrated in Fig.~\ref{matreelatest}. The algorithm terminates further clustering when the node level is more than $n$.
\begin{algorithm}
\caption{AT($D, n, L$)}
\label{BATAlgo}
$v\leftarrow$ {\tt new Node()}\\
$v.level\leftarrow L$\\
$v.cluster\leftarrow D$\\
\If{($ L \leq n$)}{
$(D_L,D_R) \leftarrow$MSMVMCA(D,2) \Comment{\textbf{$D_L,D_R$ are two clusters generated by clustering function} }
$v.left\leftarrow AT(D_L,n, L+1$)\\
$v.right\leftarrow AT(D_R,n,L+1$)\\
}
\Return $v$
\end{algorithm}
Algorithm \ref{anomalydetector} finds the leaf nodes from the \acrshort{at} and applies \acrshort{ecblof} on all the leaf nodes. It applies anomaly ranking on the data points by sorting the dataset in descending order based on \acrshort{ecblof} score.
\begin{algorithm}
\caption{MSMBTAI($D,n$)}
\label{anomalydetector}
\KwResult{Probable anomaly data-points}
$L\leftarrow 0$ \\
$AT\_root\leftarrow$AT(D, L, n)\\
$leaf\_nodes\leftarrow$find\_leaves($AT\_root$)\\
anomaly\_score $\leftarrow$ ECBLOF(leaf-nodes)\\
anomaly\_score $\leftarrow$ descending-sort(anomaly-score)\\
\Comment{This function sorts the data-instances based on anomaly scores in descending order. Data-instances with higher anomaly scores take the top positions.}

\Return Top$_N$(anomaly-score)
\end{algorithm}

\section{Data and Experiments}

\begin{table*}
\centering
\caption{A sample of anomalous exoplanets and statistics for the complete anomalous set for Dataset~D1 including \textsl{Surface Temperature} and \textsl{Surface Flux}. In total, 51 anomalies are detected out of 1682 exoplanets.}
\label{table: D1_result_ST}
\resizebox{0.8\textwidth}{!}{
\begin{tabular}{|l|l|r|r|r|r|r|}
\hline
Planet & Class & 
\multicolumn{1}{l|}{LOF} & 
\multicolumn{1}{l|}{\begin{tabular}[c]{@{}l@{}}K-NN-\\ ECBLOF\end{tabular}} & \multicolumn{1}{l|}{HBOS} & \multicolumn{1}{l|}{iForest} & \multicolumn{1}{l|}{MSMBTAI} \\ \hline
KIC-5522786 b        & thermoplanet                                  & 16                                                                  & 13                                   & 12                                 & 11                                    & 14                                 \\ \hline
LHS 1140 b           & psychroplanet                                  & 535                                                               & 93                                   & 38                                 & 57                                    & 21                                 \\ \hline
Proxima Cen b        & psychroplanet                                  & 273                                                               & 56                                   & 141                                & 111                                   & 23                                 \\ \hline
TRAPPIST-1 d         & mesoplanet                                   & 201                                                                & 45                                   & 9                                  & 63                                    & 42                                 \\ \hline
TRAPPIST-1 e         & psychroplanet                                 & 218                                                                 & 44                                   & 20                                 & 65                                    & 44                                 \\ \hline
TRAPPIST-1 f         & psychroplanet                                 & 215                                                                 & 40                                   & 19                                 & 64                                    & 50                                 \\ \hline
tau Cet e            & mesoplanet                                     & 442                                                              & 248                                  & 80                                 & 43                                    & 67                                 \\ \hline
K2-18 b     & mesoplanet                                    & 366                                                      & 283                                  & 336                                & 286                                   & 171                                \\ \hline
Kepler-61 b & mesoplanet                                    & 445                                                       & 384                                  & 260                                & 486                                   & 538                                \\ \hline
\textbf{Min}    &                & 16 & 3          & 9      & 11     & 14              \\ \hline
\textbf{Max}    &      & 1485           & 816    & 1334   & 763    & 664            \\ \hline
\textbf{Mean}   &            & 440.14       & 263.39 & 335.80 & 273.94 & 215.51 \\ \hline
\textbf{StdDev} &   & 396.84       & 176.73 & 329.34 & 180.20 & 175.60         \\ \hline
\textbf{Range}  &   & 16--1485      & 13--816 & 9--1334 & 11--763 & 14--664         \\ \hline
\end{tabular}
}
\end{table*}

\begin{table*}
\centering
\caption{A sample of anomalous exoplanets and statistics for the complete anomalous set for Dataset D1 excluding \textsl{Surface Temperature} and \textsl{Surface Flux}. In total, 51 anomalies are detected out of 1682 exoplanets.}
\label{table: D1_result}
\resizebox{0.8\textwidth}{!}{
\begin{tabular}{|l|l|r|r|r|r|r|}
\hline
Planet & Class & 
\multicolumn{1}{l|}{LOF} & 
\multicolumn{1}{l|}{\begin{tabular}[c]{@{}l@{}}K-NN-\\ ECBLOF\end{tabular}} & \multicolumn{1}{l|}{HBOS} & \multicolumn{1}{l|}{iForest} & \multicolumn{1}{l|}{MSMBTAI} \\ \hline
KIC-5522786 b   & thermoplanet                                   & 12                                                                  & 13                                   & 37                                 & 13                                    & 12                                 \\ \hline
Proxima Cen b   & psychroplanet                                  & 288                                                               & 43                                   & 85                                 & 83                                    & 17                                 \\ \hline
LHS 1140 b      & psychroplanet                                 & 415                                                                & 84                                   & 125                                & 50                                    & 21                                 \\ \hline
TRAPPIST-1 d    & mesoplanet                                   & 200                                                                 & 45                                   & 9                                  & 42                                    & 37                                 \\ \hline
TRAPPIST-1 e    & psychroplanet                                 & 203                                                                 & 48                                   & 18                                 & 49                                    & 73                                 \\ \hline
TRAPPIST-1 f    & psychroplanet                                 & 205                                                                 & 44                                   & 64                                 & 61                                    & 44                                 \\ \hline
tau Cet e       & mesoplanet                                     & 477                                                               & 203                                  & 107                                & 47                                    & 68                                 \\ \hline
K2-18 b         & mesoplanet                                    & 487                                                               & 247                                  & 176                                & 294                                   & 206                                \\ \hline
Kepler-61 b     & mesoplanet                                    & 444                                                                & 373                                  & 210                                & 461                                   & 563                                \\ \hline
\textbf{Min}    &                                                & 12                                                          & 13                                   & 9                                  & 13                                    & 12                                 \\ \hline
\textbf{Max}    &                                      & 1605                                                              & 838                                  & 646                                & 641                                  & 644                                \\ \hline
\textbf{Mean}   &        & 412.82       & 257.24        & 262.33      & 247.33    & 205.90    \\ 
\hline
\textbf{StdDev} &       & 347.29    & 188.11              & 147.05        & 138.68        & 161.93     \\ 
\hline
\textbf{Range}  &    & 12--1605      & 13--838    & 9-646  & 13--41 & 12--644    \\ 
\hline
\end{tabular}
}
\end{table*}

\noindent \textbf{Dataset:}

PHL-EC catalog serves as one of the most comprehensive datasets comprising both measured and derived features for exoplanets and their host stars. The catalog provides 68 features for the confirmed exoplanets: 13 categorical features and 55 continuous features, however, not every entry has all the features. Sometimes there are missing values for \textsl{Eccentricity} or \textsl{Surface Temperature}, for example. In the catalog, all planets are sorted into five categories based on their thermal surface characteristics: non-habitable, and potentially habitable: psychroplanet, mesoplanet, thermoplanet and hypopsychroplanet\footnote{\!\!phl.upr.edu/library/notes/athermalplanetaryhabitabilityclassification\\forexoplanets}. For reasons stated in Section~\ref{sec:2.2}, we analyse the contribution of the surface temperature as a feature in an unsupervised setting in the experiments presented below, and only consider planets whose surface temperatures are measured, and not estimated. 

This paper presents results based on two versions of this dataset (D1 and D2), both of which are described below. It is important to note that in both of the experiments with D1 and D2, features serving as measured markers for habitability, such as \textsl{Star's Habitable Zone (HZ)} and \textsl{Earth Similarity Index (ESI)} were removed. (See Appendix \ref{appenix-features} for the features from D1 and D2 used in the experiments.)  \\
\noindent \textbf{Dataset D1:}
The 2018 iteration of the PHL-EC dataset, containing \textsl{Surface Temperature} and \textsl{Stellar Flux} parameters, classifies 1681 planets into 5 habitable classes based on their surface temperatures, out of which 1631 are non-habitable. The disadvantage of class imbalance in the dataset (non-habitable: 1631, mesoplanets: 30, psychroplanets: 16, thermoplanets: 3, hypopsychroplanets: 2) hinders the robustness of any classifier in the absence of surface temperature. To compare the effect of \textsl{Surface Temperature} and \textsl{Stellar Flux}, the results include two sets of inferences from this data set, one including the above mentioned features and one without it, after the class labels are removed. 

\noindent \textbf{Dataset D2:} The latest version contains 4048 exoplanets and classifies planets into three classes, non-habitable planets constituting a majority (3993). The current version does not include \textsl{Surface Temperature} and even such important characteristics as \textsl{Gravity, Density}, or \textsl{Escape Velocity}, are available for only 706 out of 4048 of these exoplanets. Hence, only one set of inferences covering these 706 exoplanets is drawn, out of which only 6 intersect with the optimistic/conservative list of exoplanets in the PHL-HEC. Only 93 of the 706 exoplanets intersect with D1 dataset.

\begin{table*}
\centering
\caption{A sample of anomalous exoplanets and statistics for the complete anomalous set for Dataset D2 containing 6 anomalies in 706 exoplanets.}
\label{table: D2_result}
\begin{tabular}{|l|r|r|r|r|r|r|r|}
\hline
Planet & 
\multicolumn{1}{l|}{LOF} & 
\multicolumn{1}{l|}{\begin{tabular}[c]{@{}l@{}}K-NN-\\ ECBLOF\end{tabular}} & \multicolumn{1}{l|}{HBOS} & \multicolumn{1}{l|}{iForest} & \multicolumn{1}{l|}{MSMBTAI} \\ \hline
K2-18 b                                        & 174                                                                 & 134                                  & 213                                & 132                                   & 46                                 \\ \hline
TRAPPIST-1 f                                   & 70                                                                  & 67                                   & 105                                & 69                                    & 54                                 \\ \hline
TRAPPIST-1 g                                   & 73                                                                  & 71                                   & 107                                & 72                                    & 67                                 \\ \hline
LHS 1140 b                                      & 156                                                                 & 52                                   & 160                                & 55                                    & 68                                 \\ \hline
TRAPPIST-1 d                                   & 67                                                                   & 65                                   & 108                                & 65                                    & 81                                 \\ \hline
TRAPPIST-1 e                                   & 68                                                                  & 66                                   & 104                                & 73                                    & 85                                 \\ \hline
\textbf{Min}   & 67       & 52     & 104      & 55     & 46  \\ 
\hline
\textbf{Max}     & 174     & 134       & 213      & 132    & 85     \\ 
\hline
\textbf{Mean}       & 101.33      & 75.83       & 132.83   & 77.67      & 66.83   \\ 
\hline
\textbf{StdDev}    & 49.69     & 29.21    & 44.84     & 27.41   & 15.04  \\ 
\hline
\textbf{Range}    & 67--174    & 52--134     & 104--213    &55--132  & 46--85  \\ 
\hline
\end{tabular}
\end{table*}

\noindent \textbf{Preprocessing:}

In the Dataset D1, meso, thermo, psychro, and hypopsychro exoplanets are marked $1$ (anomalies) and all the non-habitable planets are marked $0$. Out of all the 1682 planets, any feature having data for less than 1600 planets is removed. Features not affecting habitability in any way as well as highly correlated features are also removed. 10 features are retained after Principal Component Analysis for both datasets to ensure fairness. For the Dataset D2, both conservative and optimistic habitable planets are considered as anomalies (marked 1) and all others are considered as normal instances (marked 0). These markers are not used only for cross-matching purposes. Only 706 of the exoplanets having non-null/non-missing values for important characteristics, such as \textsl{Density} and \textsl{Gravity}, are considered. Features irrelevant for habitability (\textsl{Date of Discovery}, etc.) as well as categorical features are removed. Finally, highly correlated features are removed.

\section{RESULTS:} The proposed solution \acrshort{msmbtai} is compared with \acrfull{lof} \citep{10.1145/335191.335437}, \acrshort{ecblof}, HBOS \citep{inproceedings}, and \acrshort{iforest} \citep{iforest}. \acrshort{lof}, and \acrshort{ecblof} are clustering based anomaly detection algorithms and these algorithms rely on a solution to identify anomalies. The solution applied is K-NN as an auxiliary tool in these approaches to identify anomalies\footnote{See Appendix \ref{appendix:methodologies} for details about these methods}. The results will be discussed in two parts, by comparing \acrshort{msmbtai} and other clustering algorithms, and secondly discussing the effect of exclusion of surface temperature and stellar flux parameters. Since the range of anomaly scores can vary between algorithms, the results are discussed in terms of anomaly rank: a ranking based on the anomaly score obtained from an algorithm, with the lower anomaly rank indicating higher chance of a sample being anomalous, i.e. habitable. Consider the following definitions used in summarizing the results in the table:
\begin{itemize}
    \item \textbf{Min:} Minimum anomaly rank among all anomalous planets
    \item \textbf{Max:} Maximum anomaly rank among all anomalous planets
    \item \textbf{Mean:} Average anomaly rank among all anomalous planets
    \item \textbf{StdDev:} Standard deviation of anomaly ranks among all anomalous planets
    \item \textbf{Range:} Maximum-Minimum anomaly rank of all anomalous planets
\end{itemize}
In Table~\ref{table: D1_result_ST}, \acrshort{msmbtai} produces better result in comparison to other algorithms in multiple parameters. iForest achieves the lowest maximum rank and  \acrshort{msmbtai} achieves the minimum mean rank in Table \ref{table: D1_result}. \acrshort{msmbtai}, among all algorithms, also assigns the lowest rank to such promising potential habitable exoplanets as Proxima Cen~b and the TRAPPIST-1 planets.  

\noindent 
\textsl{Surface Temperature} and \textsl{Stellar Flux} serve as hard markers for classification of exoplanets as habitable. Hence, after removing these features, the algorithms are expected to perform significantly worse. However, comparing Table \ref{table: D1_result_ST} and Table \ref{table: D1_result}, most algorithms perform very similar or even better after excluding parameters of surface temperature and stellar flux. We note that \acrshort{msmbtai} assigns lowest rank to Proxima~Cen~b, tau~Cet~e and TRAPPIST-1 exoplanets. The nominal differences between the two tables indicate that sufficient clustering information can be obtained from the observed features alone, and fine-tuned unsupervised methods can be used in the future to indicate habitability of newly observed exoplanets. The results are especially motivating for the scientific community as features indicating habitability, such as minimum and maximum habitable zone of a star and similarity indices, like ESI, are not included in either of the computations.

\begin{table*}
\centering
\caption{Description of the parameters of employed algorithms and their values for exoplanet datasets.}
\label{table:all_params_desc}
\begin{tabular}{|l|l|c|}
\hline
\textbf{Algorithm} & \textbf{Parameter Description} & Values in \textbf{D1, D2}\\
\hline
MSMBTAI($n$)& $n$: the maximum tree level & (4) \\
\hline
MSMVMCA   &$v$: the number of versions in the population, &  \\
($v,s_1,s_2,n_1,n_2,F_t$) & $s_1$, $s_2$: proportions of stage1 and stage2 & ($12,0.5,0.5,10,10,0.5$) \\
& populations becoming parent,  & \\
& $n_1$, $n_2$: max iterations number of stage1 and stage2, &   \\
& $F_t$: the fitness threshold.  &  \\
 \hline
LOF$(k)$ & $k$: number of neighbours included in density computation &  (Average over 10--25)\\
 \hline
HBOS$(bins,tolerance)$ & $bins$: number of histogram bins to form, & \\
& $tolerance$: determines the flexibility while &  (10,0.5)   \\
& dealing with samples lying outside the bins.  &  \\
 \hline
iForest$(estimators,samples)$ & $estimators$: number of trees in the ensemble, &  \\ 
& $samples$: number of samples drawn to train & (100,256)  \\
& each base estimator. &  \\
\hline
\end{tabular}
\end{table*}

\section{Conclusion}

Earth-like rocky planets could amount to as many as 6 billion in the Milky Way \citep{{6billion},{latestRockyplanets}}. Numerous space missions, current and planned for near future, are searching for potentially-habitable planets, for the possibility of exolife -- life elsewhere in the Universe. Humanity is also looking for the second `Earth' -- a planet habitable for us, but preferably uninhabited, so that colonization opportunities, initially postulated by \citet{https://doi.org/10.1111/j.1465-7295.2009.00225.x} and recently considered through a quantitative model \citep{experthabitat}, can be explored. Inferences in astronomy do not rely absolutely on machine learning methods but benefit a great deal from such methods. It is the combination of ML-based inferences and domain knowledge that help reduce the search for habitable planets significantly due to the proposed method. Therefore, the outcome of MSMVMCA-based Anomaly detection is promising and relevant for the scientific community. The results from the clustering based algorithm were cross-matched with both optimistic and conservative lists of PHL-HEC habitable exoplanets. There are 42 matches out of 51 in the D1 set, and all 6 anomalies detected in D2 set match with PHL-HEC potentially habitable planets. Out of 6 discovered anomalies in D2, TRAPPIST-1~g planet is the only mismatch with D1. This means that Trappist 1-g is not detected as an anomaly in  D1 dataset, but is detected as an anomaly when MSMA is applied on D2 dataset. This is an interesting observation, as Trappist 1-g is may be too cold to host life -- its equilibrium temperature is 198.6 K (-75°C) (e.g. \citet{Wolf}). In fact, it is deemed not habitable by the Solar Equivalent Astronomical Unit (SEAU) criteria \citep{ExoKyoto}. The anomalous instances are cross-validated with domain knowledge-based expectations of habitable candidates and not by commonly used metrics, as the class labels in the PHL-EC are not beyond reasonable doubt. We have also shown the  efficacy of our clustering method on additional datasets. As future work, the following will have to be considered: 
\begin{enumerate}
\item It is required to improve the computation cost of the \acrshort{msmvmca} in comparison to other clustering algorithms.
\item \acrshort{msmvmca} has established itself as a clustering algorithm and been applied in detection of anomalous exoplanets. However, clustering algorithms such as K-means, K-medoids have been employed extensively in various domains, and it is required to implement \acrshort{msmvmca} in other research domains as well to emphasize its strength.
\end{enumerate}

\section*{Acknowledgements}

The authors would like to thank the Science and Engineering research Board (SERB), Department of Science and Technology, Government of India, for supporting our research by providing us with resources to conduct our experiments. The project reference number is: EMR/2016/005687. MS acknowledges the financial support by the Department of Science and Technology (DST), 
Government of India, under the Women Scientist Scheme (PH), project reference number SR/WOS-A/PM-17/2019.
 
\section*{Data Availability}

Data and the codes associated with this manuscript are available on several websites:\\
1. CDHS/CEESSA Datasets and Catalogs:
AstrIRG (Astroinformatics Research Group) website, http://astrirg.org/projects.html\\
2. Codes to analyze exoplanetary data from PHL's HEC catalog:  https://github.com/SuryodayBasak/Exoplanets\\Analysis/tree/master/MLAnalysis



\appendix


\section{MSMVMC Qualify as a Clustering Algorithm}\label{appen-msmvmca}

Before applying on the exoplanet data, it was crucial to verify the effectiveness of the \acrshort{msmvmca} as a clustering algorithm by applying it on various benchmark datasets. \acrshort{msmvmca} has been applied on Iris, Glass, Seed, Knowledge, Libras, and Sonar datasets \citep{iris}. These are benchmark data sets used extensively in machine classification (see the detailed description of each dataset in subsection~ A.1). \acrshort{msmvmca} has been also compared with other standard clustering algorithms such as K-means, K-medoids in terms of Rand and Jaccard indices -- two metrics widely used in the performance comparison.
\begin{itemize}
    \item Rand Index:
    A Rand index is used to measure the similarity between two data clusterings. The Rand index is related to the clustering accuracy though it can be applied when class labels are not available. The range of the Rand index is 0 to 1. Higher the value, better the accuracy of the data clustering.
    \item Jaccard Index: 
    A Jaccard index is a statistical tool used to measure the similarity and diversity of sets. It is also known as a Jaccard coefficient. It is used to compare the similarity between finite sets. It is defined as the size of the intersection divided by the size of the union of the comparing sets. The range is from 0\% to 100\%. The higher the percentage, the more similarity between the two datasets.
\end{itemize}
Two different types of fitness functions are employed in stages. The first stage has used Silhouette, and the second stage has implemented Davis-Bouldin (DB) as a fitness function (See Appendix~\ref{appendix:fitness} for more details on the fitness functions). A higher Silhouette value means it has achieved better clustering, whereas a smaller DB value indicates a better result. We have assumed that the data points which belong to the same clusters are co-located together. For example, a dataset of 10 data points, where the first 2 data points belong to clusters1, the next 5 data points belong to cluster 3, and the last 3 data points are part of cluster 2. The order looks as follows:  1133333222. The result has been demonstrated in Table~\ref{table:msmvmca}, and it is evident from the table that \acrshort{msmvmca} has outperformed other clustering algorithms.

\begin{table*}
\caption{\acrshort{msmvmca} performance on benchmark data: \acrshort{msmvmca} outperformed other algorithms in terms of both Jaccard and Rand coefficients.}
\label{table:msmvmca} 
\begin{tabular}{|l|c|c|c|c|c|c|c|}
\hline
   Dataset&\multicolumn{2}{c|}{MSMVMCA}& \multicolumn{2}{c|}{K-means}&\multicolumn{2}{c|}{K-medoids} \\
   \hline
   &Rand& Jaccard&Rand&Jaccard&Rand&Jaccard\\
 \hline
 Iris&0.9186& 0.9333& 0.8797& 0.44& 0.8922& 0.4266\\
 \hline
 Glass&0.7665&0.2663& 0.6718& 0.2429& 0.7302& 0.1869 \\
 \hline
    Seed&0.9566 &0.3380&0.8743& 0.0142& 0.8713& 0.2714 \\
\hline
    Knowledge &0.8099& 0.6551&0.7590& 0.2068& 0.7181& 0.5241 \\
\hline
 Libras& 0.9125&0.1166&0.8911& 0.1083& 0.8748& 0.1 \\
\hline
 Sonar&0.5094 &0.5769&0.5032& 0.4471& 0.5054& 0.5625 \\
\hline
\end{tabular}
\end{table*}

\begin{figure*}
\centering
\includegraphics[width=0.7\textwidth]{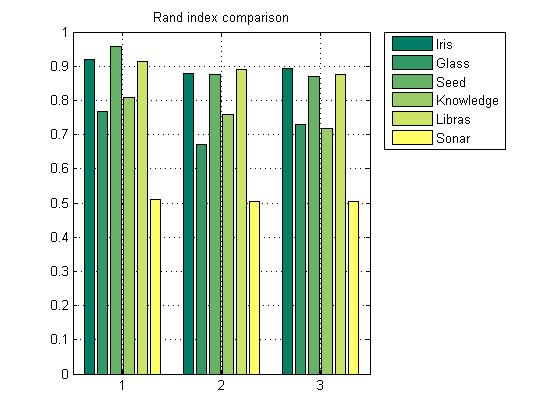}
\caption{Comparison of \acrshort{msmvmca} performance with K-means and K-medoids. The $X$-axis is the algorithm, where 1, 2 and 3 represent \acrshort{msmvmca}, K-means and K-medoids, respectively. The  $Y$-axis represents the Rand index. It is evident that \acrshort{msmvmca} has achieved a higher Rand index in comparison to other algorithms.}
\label{rand}
\end{figure*}

\begin{figure*}
\includegraphics[width=0.7\textwidth]{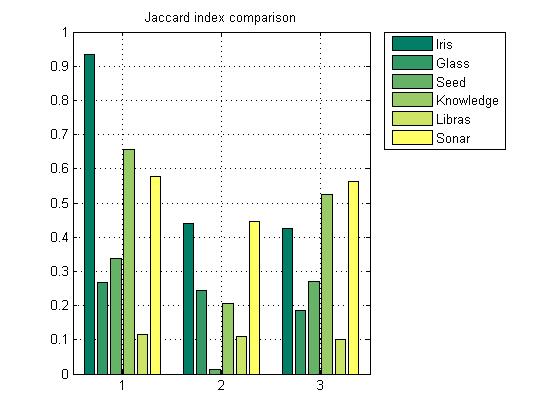}
\caption{Comparison of \acrshort{msmvmca} Jaccard index with K-means and K-medoids. Here, 1, 2 and 3 in $X$-axis denote MSMVMCA, K-means and K-medoids, respectively. $Y$-axis is the Jaccard index. \acrshort{msmvmca} has attained a better result in comparison to other standard algorithms.}
\label{jaccard}
\centering
\end{figure*}

\subsection{Description of Datasets}

\subsubsection{Iris}

Iris is a well-known dataset used widely for clustering purposes in the field of pattern recognition. It has a total of 150 data points and 4 attributes. The total of three classes, each containing 50 data instances, were used for the task of anomaly detection. It is evident from Table~\ref{table:msmvmca} that \acrshort{msmvmca} has achieved higher accuracy in both the Rand as well as Jaccard indices. For example, \acrshort{msmvmca} has achieved Rand index 0.91, whereas K-means achieves 0.87 and K-medoid achieves 0.89. \acrshort{msmvmca} has considered 100 versions in the initial population, where every chromosome, or version, was generated randomly. As we have assumed that the data points belonging to the same clusters are co-located, the initial population is generated in the same fashion. For example, version 1 randomly decided that the first 10 data instances are cluster 1, the next 100 belong to cluster 2, and the rest of the data instances are part of cluster 3. The same methodology has been applied to all other datasets to maintain uniformity. For both stages, Davis-Bouldin was used as a fitness function. However, the best version is selected from the population based on the Silhouette score.

\subsubsection{Seed}
This dataset comprises kernels belonging to three different types of wheat, with total of 210 data instances and 7 attributes. The dataset has three clusters, each having 70 data points. The \acrshort{msmvmca} considered 100 version and Silhouette and DB as fitness functions in stage 1 and stage 2, respectively.
\acrshort{msmvmca} has attained 0.95 Rand index, which is higher than 0.87 achieved by both K-means and K-medoids.

\subsubsection{Sonar}

This sonar dataset consists of 208 records and 60 attributes. All these data instances belong to two classes, either Rock or Metal. Out of all these records, 111 data points were obtained by bouncing sonar signals off a metal cylinder at various angles and under various conditions. 97 records have been obtained from rocks under similar conditions. It is apparent from Table \ref{table:msmvmca} that our proposed algorithm has outperformed K-means, K-medoids in terms of both Rand and Jaccard indices.  

\subsubsection{Libras}

Libras comprises 360 data instances, and the total number of attributes is 91, which is higher than previous datasets. It contains 15 classes, each with 24 data points, and each class is a reference to a hand movement type in Libras. \acrshort{msmvmca} attains higher accuracy in both Rand and Jaccard indices.

\subsubsection{Knowledge}

This dataset has 403 records, and the total number of attributes is 5. It has a total of 4 classes (Very Low, Low, Middle, High). Very Low has 50 instances, Low has 129 instances, Middle and High have 122 and 130 instances, respectively. 
\acrshort{msmvmca} reached 0.80 in the Rand index, which is almost 0.09 more than K-medoids, and in the case of Jaccard index, the \acrshort{msmvmca} has three times higher value than K-means.

\subsubsection{Glass}

The Glass dataset contains a total of 214 instances, and the total number of attributes is 10. This dataset is used extensively by many researchers in their papers. The same algorithms have been applied to the Glass dataset to evaluate the performance of the K-means algorithm after the dimension reduction.
The types of glass present in the aforementioned dataset are
\begin{enumerate}
\item  building\_windows\_float\_processed 
\item building\_windows\_non\_float\_processed  
\item vehicle\_windows\_float\_processed 
\item containers  
\item tableware 
\item headlamps
\end{enumerate}
\acrshort{msmvmca} has performed better than other well-known algorithms as is evident from Table~\ref{table:msmvmca}.
In Figures 
\ref{rand} and \ref{jaccard},
\acrshort{msmvmca}, K-means, and K-medoids are denoted as 1, 2, and 3, respectively.

\subsection{Setting Parameters for Different Datasets}

The parameter settings for K-medoids and K-means algorithms for different datasets are shown  in Table~\ref{table:params1}. \acrshort{msmvmca}  algorithm had the same parameter setup for all datasets: MSMVMCA($v,s_1,s_2,n_1,n_2, F_t$)=(100,0.5,0.5,10,10,0.7). 

\begin{table}
\caption{Parameter settings on different datasets}
\label{table:params1}
\begin{tabular}{l c c c c l l}
\hline
Algorithm & Iris & Seed & Sonar & Knowledge & Glass & Libras \\
 \hline
K-medoids ($k$) & (3) & (3) & (2)  &  (4) & (6) & (15)\\
\hline
K-means ($k$) & (3) & (3) & (2) & (4) & (6)& (15)\\
\hline
\end{tabular}
\end{table}

\section{List of features used in clustering Data sets D1 and D2 for anomalous (habitable) candidate detection}
\label{appenix-features}

\noindent {Features used for clustering for Dataset D1 (Features indicating (*) are included in one experiment and excluded in the other):} 
\textsl{Planet Mass, *Planet Stellar Flux (Mean), Star Size from Planet, Planet Radius, *Planet Surface Temperature (Mean), Star Temperature Effective, Planet Density, Planet Surface Pressure, Star Luminosity, Planet Gravity, Planet Period, Star Right Ascension, 
Planet Escape Velocity, Star Mass, Star Declination,
Planet Semi Major Axis, Star Radius} \& \textsl{Star Magnitude from Planet}.
\vskip 0.1in
\noindent {Features used in \acrshort{msmvmca} and other clustering techniques for Dataset D2:} \\
\textsl{Planet Mass, Star Radius, Planet Density, Planet Period, Star Temperature, Planet Distance, Star Right Ascension, Planet Escape  Velocity, Planet Radius, Star Declination, Planet Potential (planet gravitational potential in earth units), Star Mass} \& \textsl{Planet Gravity}.

\section{Benchmark Methodologies: Anomaly Detection}\label{appendix:methodologies}

This section describes various anomaly detection algorithms, which were used for comparison with \acrshort{msmbtai}.

\subsection{K-NN Global anomaly detection}

For every data point, the $k$ nearest neighbors are calculated. The distance metric, e.g. L1, L2, Euclidean or Mahalanobis, can be decided according to the dataset. The anomaly score can be calculated according to measures described below:
\begin{enumerate}
\item largest: use the distance to the $k$-th neighbor as the outlier score
\item mean: use the average of all $k$ neighbors as the outlier score
\item median: use the median of the distance to $k$ neighbors as the outlier score
\end{enumerate}

The anomaly score, however, is heavily dependent on normalization criteria and the number of dimensions. Another challenge is to be able to decide on the value of `$k$'. While in supervised algorithms, $k$ can be determined with cross-validation, deciding on a $k$ in unsupervised cases is more challenging. Hence, generally, an average over many values of $k$ is used to make a fair comparison with the rest of the algorithms. The algorithm performs well for global anomalies since for a data point to have a high score, it should be away from all the clusters in the dataset.

\subsection{Local Outlier Factor (LOF)}

As the name suggests, LOF is a local outlier scoring method that can be used for global anomalies as well. It is a density-based algorithm that relies on the $k$ nearest neighbors. The estimated local density of a point is the number of its neighbors divided by the cumulative sum of distances to its neighbors. LOF is then calculated by dividing the average density of the neighbors by the point's density. 
Suppose $N(p)$ is the set of neighbors of point $p$, $k$ is the number of points in this set, and $d(p,x)$ is the distance between points $p$ and $x$. The estimated density is:
\be
\hat{f}(p) = \frac{k}{\Sigma_{x\in N(p)} d(p,x)}\,,
\ee
and the local outlier factor score is:
\be
LOF(p) = \frac{\frac{1}{k}\Sigma_{x\in N(p)}\hat{f}(x)}{\hat{f}(p)}\,.
\ee
Since LOF is a ratio of densities, normal instances would get a value close to 1, whereas outliers will get larger values. Here again, the value of $k$ needs to be adjusted.

\subsection{Histogram-based Outlier Score (HBOS)}

HBOS is a statistical algorithm, which calculates the anomaly score by creating a univariate histogram for each feature of the dataset, assuming independence of the features. The disadvantage of assuming feature independence becomes less severe when the dataset has a high number of dimensions, which is true for the case of the PHL-EC dataset. The density estimate is represented by the height of each bin of the histogram. For each feature, the histogram is normalized to $[0,1]$ and HBOS for each instance $v$ is computed as a product of the inverse of the estimated density:
\be
HBOS(v) = \sum\limits_{i=0}^d \log{\left (\frac{1}{hist_i(v)}\right)}\,,
\ee
where $d$ is the number of dimensions, $v$ is the vector of features, and $hist_i(v)$ is the density estimate of each feature. Inverting the score ensures anomalies have a higher score than normal instances. 

\subsection{Isolation Forest}

Isolation Forest is an ensemble method which 'isolates' data points by randomly selecting a feature and then randomly selecting a split value to divide the data points into 2 nodes. Recursive partitioning will result in each observation residing in a leaf node, and the number of splittings required to isolate a sample is equal to path length from root to the leaf of the tree. If this length is averaged over a forest of many such random trees, it will act as a measure of the outlying behavior, producing shorter length for anomalies on average.

\section{Fitness Functions}\label{appendix:fitness}

\subsection{Silhouette Index}

The Silhouette index refers to the interpretation and validation of the consistency within the clusters of data. It measures how similar an object to its cluster (cohesion) in comparison to another cluster (separation). The ranges of the Silhouette vary from $-1$ to $+1$. A higher Silhouette indicates the object is matched perfectly to its cluster and poorly to other clusters. The value $+1$ means the sample is far away from the neighboring cluster and very close the cluster where it is assigned. $-1$ means the sample is near to the neighboring clusters and it is not so close to the assigned cluster. $0$ indicates that it is at the boundary of the distance two clusters. $+1$ is the ideal value and $-1$ is the least preferable. The higher the value, the better the clustering of the dataset. Assume the data has been partitioned into $K$ clusters using some standard clustering mechanism. Now, $i$ is a data-point and $a(i)$ is the average distance of all the data-points belong to the same cluster. $b(i)$ is the mean distance of the data point $i$ to all other data points of a neighboring cluster $(B)$ to which $i$ is not a member. And the neighbor cluster to which the data-point $i$ is closest in comparison to other clusters. The Silhouette can be calculated as follows,
\begin{equation}
    s(i)=\frac{b(i)-a(i)}{max(b(i),a(i))}\,.
\end{equation}
$s(i)$ will become $1$ if the $b(i)$ value is very large and $a(i)$ is very small: $b(i)>>a(i)$. This situation may happen when $a(i)$ is very close to other data points inside the cluster, and far away from the closest neighbouring cluster. This metric is used as a fitness function in stage1 in \acrshort{msmvmca}.

\subsection{Davis-Bouldin}

Davis-Bouldin is a metric widely used for evaluating the clustering algorithm. It is the ratio between the within-cluster distances and the between-cluster distances. In the same way, it computes the average of the overall clusters and it is very easy to implement. The lower the score better the clustering \citep{Davies}. The lowest value of the Davis-Bouldin that can be achieved is $0$. We have considered this metric as a fitness function in stage 2 in the proposed algorithm.


\label{lastpage}
\end{document}